# A Jump Distance-based Bayesian analysis method to unveil fine single molecule transport features

Sylvain Tollis[1,2]

[1]: Wellcome Trust Centre for Cell Biology, University of Edinburgh, Michael Swann Building, Max Born Crescent, Edinburgh, EH9 3BF, United Kingdom

[2]: Institut de Biologie et Génétique Cellulaires, CNRS UMR5095, 1 rue Camille Saint-Saën - CS61390, Université de Bordeaux - Campus Carreire, 33077 Bordeaux Cedex, France

Correspondence : s.simyion@wanadoo.fr

Running title: Jump Distance-based Bayesian method

## Abstract

Single-molecule tracking (SMT) methods are under considerable expansion in many fields of cell biology, as the dynamics of cellular components in biological mechanisms becomes increasingly relevant. Despite the development of SMT technologies, it is still difficult to reconcile a sparse signal at all times (required to distinguish single molecules) with long individual trajectories, within confined regions of the cell and given experimental limitations. This strongly reduces the performance of current data analysis methods in extracting meaningful transport features from single molecules trajectories. In this work, we develop and implement a new mathematical analysis method of SMT data, which takes advantage of the large number of (short) trajectories that are typically obtained with cellular systems *in vivo*. The method is based on the fitting of the jump distance distribution, e.g. the distribution that represents how far molecules travel in a set time interval; it uses a Bayesian approach to compare plausible molecule motion models and extract both qualitative and quantitative information. Finally, the method is tested on *in silico* trajectories simulated using Monte Carlo algorithms, and ranges of parameters for which the method yields accurate results are determined.

# Introduction

Recent improvements in both the spatial and temporal resolution of fluorescence microscopy has allowed to detect and track individual biological molecules over time in living cells and tissues. Single Molecule Tracking (SMT) experiments have become increasingly popular, as illustrated by the attribution of the 2014 Nobel Prize of Chemistry to the pioneer work of Betzig, Hell and Moerner, in a variety of fields including plasma membrane dynamics (1, 2), high resolution imaging of internal organelles (3, 4) or intracellular transport (5). With the improvement of fluorophores stability (6-8) and camera sensitivity, applications of SMT have extended to fields as diverse as cellular neurosciences (9-11), biomedicine (12, 13), the dynamics of cell surface receptors (14, 15), transcription factors (16) or Calcium channels (17). While knowledge on the static and genetic aspects of interactions among biological molecules of interest builds rapidly, very little is known about the complementary dynamic behaviour of these molecules. SMT is therefore an appealing technique to provide information on their local environment and uncover the spatial and temporal regulation of individual and collective transport properties in cells (18-28).

However, inferring reliable information from single trajectories requires extracting meaningful signals from stochastic molecular transport. Many modes of biological transport are well-characterized mathematically (29), including passive diffusion through diverse media (30), polymer networks (31), active motion avoiding mobile or immobile obstacles (32), or directed motion along cytoskeletal tracks (33); however mathematical analysis tools that fully account for trajectory variability in transport model evaluation while overcoming experimental limitations such that the sampling rate, the total acquisition time, localization errors, or trajectories crossing (34-36) are still to be found (18, 37).

SMT experiments provide raw data as time series of imaged frames. Computer algorithms are generally used to detect and localize individual molecules in consecutive frames, building single molecule trajectories over time. In the past decade, several research groups have developed powerful tracking algorithms, able to deal with fluorophore blinking, focal drift, and the merging or splitting of trajectories for instance (1, 2, 37-41). However, owing to the limited optical resolution, it is still difficult to follow single molecules for a long time while keeping the occurrence of trajectories merging or clustering to a reasonably low level, except in artificial, controlled *in vitro* systems (42-44). This is especially true when tracked molecules are concentrated in a specific subcellular region (organelles, liposomes, molecule clusters …). In addition, photo-bleaching of the fluorophores introduces physical limitations to the length of acquired trajectories (41, 45). Hence, SMT strategies often yield quite short trajectories (less than 100 points) that require specific analysis methods (45, 46). Furthermore, biological molecules often experience multiple transport modes over the course of a long trajectory (47, 48). In this situation, specific analyses approaches are required to separate the different transport modes and estimate their relative contribution to individual trajectories. How can we extract the heterogeneous dynamics enclosed in a population of single molecule trajectories?

In practice, even in the simplest case where molecules experience free Brownian diffusion the extraction of biological information is often challenging (18, 36, 49). In particular, there is currently a need for a robust analysis of short trajectories obtained from SMT experiments. Despite recent works that tackled this issue (45, 46), the characterization of the underlying motion model is still broadly done through the analysis of the mean-square displacement (MSD) (10, 50-55). Free diffusion is characterized by linear MSD vs time lag plots, the slope being proportional to the diffusion coefficient $D$ (56). Sublinear plots indicate constrained diffusion and/or molecular interactions, while supralinear MSD indicates active transport processes (56, 57). The key limitation of such analysis is that the MSD shows strong statistical fluctuations for time lags representing a significant fraction of the entire trajectory; for short trajectories, this noise affects all the points of the MSD vs time lag plots, making the determination of the motion parameters almost impossible (58, 59). A minimal number of 100-1000 trajectory points seem to be generally required for meaningful MSD analysis (36, 60), which is hard to achieve experimentally. Ensemble averaging of individual MSD curves usually damps the fluctuations and improves the accuracy of the diffusion coefficient measurement (36). However, if multiple transport modes are present, ensemble-averaging tends to average out the less frequent transport modes to the benefit of the dominant mode (47, 61). Despite these limitations, MSD-based analysis of SMT data is still widely used in mammalian cells (62), yeast (54), *C. elegans* (63, 64)...

Other measurable quantities have been proposed as alternatives to the MSD (46, 65-72). Among them, the jump-distance (JD), defined as the distance travelled by a single molecule during a fixed time lag $\tau$, has a distribution (JDD) over the population of molecules that reflects fine features of the underlying transport. The JDD analysis takes advantage of the fact that modern, powerful imaging systems and tracking algorithm produce generally a large number of individual trajectories allowing to plot well-resolved JD distributions. There exist closed-form mathematical formulations of JDDs for many motion models (56, 69, 73); conveniently, the JDD for composite populations encompassing subpopulations experiencing different transport modes is a linear combination of the JDD of each mode and of the relative size of the subpopulation (see Materials and Methods, M&M). To calculate the JDD, all trajectories must be of same duration $\tau$, which can be achieved by splitting longer trajectories into sub-trajectories of duration $\tau$ that will be considered as an individual trajectory. This procedure naturally separates multiple transport modes that may be experienced by a single molecule along its entire trajectory. Then, the JDD analysis should overcome two of the major limitations of *in vivo* SMT, e.g., short trajectories, and possible multiple motion modes, to the detriment of losing single-trajectory information. Motion parameters, such that the diffusion coefficient, are therefore obtained at a population level. This method has been already used successfully in various contexts (9, 68, 69). However, fitting the JDD obtained from *in vivo* data requires the knowledge of an underlying motion model *a priori*, which is in general unknown. Then, multiple motion models can lead to satisfactorily fits of the experimental data, hence the requirement for a model selection procedure.

Model selection can be achieved through the minimization, with respect to model parameters, of the squared error between experimental data and model prediction for a given quantity. Models can be classified according to their least square difference to the experimental data, a smaller error being interpreted as a better model. This procedure generally selects overly complex models (with more parameters), which usually produce better fits, while the particular features of a given dataset might only reflect stochastic variations of a simple underlying transport mechanism (29, 36, 58, 74). Although the definition of adequate information criteria (75) or maximal likelihood estimators (45), or the use of probabilistic graphical models (76) have been proposed to overcome this issue, there is no consensus on how to extract meaningful information from inherently stochastic SMT data. Recently, approaches based on Bayesian inference have proven able to successfully discriminate between competing motion models (75, 77-80), and in handling noise and experimental limitations in other biological applications including fluorescence correlation spectroscopy (76, 81-85). However, to the best of our knowledge, in the context of SMT Bayesian methods have been mostly restricted to MSD-based analysis, and might therefore be inaccurate to analyze short trajectories and/or long trajectories encompassing multiple transport modes (18, 29, 74, 77).

In this article, we derive a Bayesian approach for motion model selection, based on the fitting of the experimental jump distance distribution with model predicted JDDs. Our procedure evaluates the likelihood of competing underlying models to have produced the observed data and estimates the optimal model parameters. We demonstrate mathematically that model likelihood computation reduces to two steps: the minimization of a generalized squared error function yielding, for each putative motion model, the optimal set of parameters, and subsequent integration of raw model probabilities over a narrow region of the model parameter space. The latter step is crucial to deal with the uncertainty on optimal parameter estimation that arises from biological noise, and was shown to favour simple models to the detriment of overly complex models (77). To estimate the accuracy of our approach, we simulate *in silico* trajectory sets using Monte Carlo methods for free diffusion (model D), anomalous diffusion (model A), noisy directed motion (model V), and composite models encompassing two subpopulations of trajectories corresponding to different transport modes (models DD, DV, DA), and analyse simulated trajectories to recover the underlying motion parameters. These tests allow estimating the uncertainties on motion parameter extraction, which are used to reduce, for each model, the parameter space over which model probabilities are integrated to yield the desired model likelihood. This procedure reduces both the effects of stochastic variations on model parameters determination and the risk of selecting overly complex models, a general property of Bayesian inference (77-80, 84-86).

The paper is organized as follows: first, we use simulated JDDs to systematically determine the accuracy of parameter determination as a function of trajectory length, number of JDD bins, number of simulated trajectories, and model parameters for all 6 models introduced above. We show that our fitting procedure is accurate for all motion models and for large parameter ranges, even for short trajectories, and we determine the parameter ranges under which the accuracy is reduced. Then, for particular sets of biological parameters representative of these ranges, we estimate the accuracy of the JDD-based Bayesian model selection procedure and show that composite populations are correctly identified in biologically-relevant situations.

## Materials and Methods

### Characterizing single molecule motion using the Jump Distance Distribution

We aim to analyze a set of $N$ individual planar trajectories, or fraction of trajectories, labelled with the index $j=1..N$. All trajectories comprise of $M+1$ points of planar Cartesian coordinates and have an identical duration $\tau = M\delta t$ with $\delta t$ the observation time step. For each trajectory, the jump distance (JD) is defined as the Euclidian (geometrical) distance between its first and last points. The jump distance distribution (JDD) is defined by classifying the $N$ jump distances in $N_b$ ordered bins of fixed size $\delta r$ (see Fig. S1A). By definition, a trajectory is added to the $i^{th}$ bin if its JD satisfies $(i-1)\delta r \leq JD < i\delta r$. The JDD is therefore defined by the set of integer numbers $\left\{ N_i, i=1..N_b, \sum_{i=1}^{N_b} N_i = N \right\}$ corresponding to the number of trajectories in each bin for the analyzed dataset.

For the simple models D, V and A considered in this study, previous mathematical studies have derived the theoretical jump distance probability distributions (56, 73, 87-89). The corresponding jump distance distributions for the simple models $\left\{ \tilde{N}_i^{D,V,A}, i=1..N_b \right\}$ are easily obtained by multiplying the probabilities by the number of molecules, and we get respectively:

$$\tilde{N}_i^D = N\delta r \frac{r_i}{2D\tau} e^{-\frac{r_i^2}{4D\tau}}, \tilde{N}_i^V = N\delta r \frac{r_i}{Mk_V} e^{-\frac{r_i^2 + V^2\tau^2}{4D\tau}} I_0\left(\frac{\tau V r_i}{Mk_V}\right)$$

where $r_i = (i-1/2)\delta r$ and the averaging over the isotropically distributed directed tracks has been performed, $Mk_V$ is the variance of the molecule position over the $M$ supposedly independent steps of a trajectory, and $I_0$ is the modified Bessel function of the first kind. $\tilde{N}_i^A$ is defined by an inverse Laplace transform:

$$\tilde{N}_i^A = N\delta r \frac{r_i}{D_\alpha} \int_{-i\gamma-\infty}^{-i\gamma+\infty} e^{ipr} \frac{(ip)^{\alpha-1}}{2\pi} K_0\left(\frac{r_i}{\sqrt{D_\alpha}}(ip)^{\alpha/2}\right) dp$$

where $K_0$ is the modified Bessel function of the second kind, and the cutoff $\gamma$ is small. For numerical evaluations of $\tilde{N}_i^A$ the cutoff was set to the working precision of our computations, and the integral boundaries were set to $\pm 300$ for $\alpha \geq 0.5$ and $\pm(300)^{0.5/\alpha}$ for $\alpha < 0.5$ to account for the slow decrease of the Bessel function for small $\alpha$.

The theoretical JDD for models encompassing two subpopulations of molecules experiencing different motion modes are linear combinations of $\tilde{N}_i^{D,V,A}$:

$$\tilde{N}_i^{DD} = f_D\tilde{N}_i^D + (1-f_D)\tilde{N}_i^{D_2}, \tilde{N}_i^{DV} = f_D\tilde{N}_i^D + (1-f_D)\tilde{N}_i^V, \tilde{N}_i^{DA} = f_D\tilde{N}_i^D + (1-f_D)\tilde{N}_i^A$$

where $f_D$ denotes the fraction of freely diffusing molecules.

**Bayesian approach for model selection on JDD data**

For $K$ competing models $M_{k=1..K}$ (here $K=6$) and an observed JDD $\{N_i\}$, the probability of each model to be the actual transport mode that produced $\{N_i\}$ is given by the Bayes theorem (77):

$$P(M_k | \{N_i\}) = \frac{P(\{N_i\}|M_k)P(M_k)}{\sum_{k=1}^K P(\{N_i\}|M_k)P(M_k)}$$

With no *a priori* knowledge on the underlying motion model, all plausible models should be considered equiprobable. Hence, $P(M_k) = 1/K$ and:

$$P(M_k | \{N_i\}) = \frac{P(\{N_i\}|M_k)}{\sum_{k=1}^K P(\{N_i\}|M_k)}$$

Additional *a priori* information can be accounted for by "preferring" such or such model, e.g. specifying particular prior probabilities $P(M_k)$. The marginal probability $P(\{N_i\}|M_k)$ is obtained by integrating over the model parameter space:

$$P(\{N_i\}|M_k) = \int P(\{N_i\}|M_k,\beta_k)P(\beta_k|M_k)d\beta_k$$

where $\beta_k$ stands for the entire parameter set.

To calculate $P(\{N_i\}|M_k,\beta_k)$, we assume that the jump distances along different trajectories are independent random variables. Given a motion model $M_k$ with its parameters $\beta_k$, for a single trajectory the probability $p_i$ that the jump distance falls in the $i^{th}$ bin is deduced from the corresponding theoretical JDD, $p_i(\beta_k) = \tilde{N}_i^{M_k}(\beta_k)/N$. Since the jump distances of different trajectories are assumed to be independent, $P(\{N_i\}|M_k,\beta_k)$ satisfies a multinomial distribution:

$$P(\{N_i\}|M_k,\beta_k) = \frac{N!}{N_1!N_2!...N_{N_b}!}(p_1)^{N_1}(p_2)^{N_2}...(p_{N_b})^{N_{N_b}}$$

with the constraint $\sum_{i=1}^{N_b} N_i = N$. Although exact, this expression is untractable in numerical computations, so we proceeded to approximate it using the Laplace's saddle-point approximation.

Starting with the Stirling formula, and using the renormalized bin populations $y_i = N_i / N$ which satisfy $\sum y_i = 1$ we get:

$$P(\{N_i\} | M_k, \beta_k) \approx e^{-N\left[\sum_{i=1}^{N_b} y_i (\ln y_i - \ln p_i) - \frac{\ln\sqrt{2\pi N}}{N} + \frac{\sum_{i=1}^{N_b} \ln\sqrt{2\pi N_i}}{N}\right]}$$

In the latter expression, the last two terms are of respective orders $(\ln N)/N$ and $N_b (\ln N)/N$, both $\ll 1$ provided that the number of bins is small compared to (large) number of trajectories (typically, several thousands). Therefore, $P(\{N_i\} | M_k, \beta_k)$ is finite for configurations $\{y_i\}$ that (almost) minimize $F(\{y_i\}) = \sum_{i=1}^{N_b} y_i (\ln y_i - \ln p_i)$, and for other configurations $P(\{N_i\} | M_k, \beta_k) \approx 0$ owing to the factor $e^{-N}$.

The minimization of $F(\{y_i\})$ with respect to $\{y_i\}$ under the constraint $G(\{y_i\}) = \sum_{i=1}^{N_b} y_i - 1 = 0$ is achieved for $\vec{\nabla} F = \varphi \vec{\nabla} G$, which yields $\forall i, 1 + \ln y_i - \ln p_i = \varphi$. From the normalization condition $\sum y_i = \sum p_i = 1$ we get the Lagrange multiplier $\varphi = 1$ and finally $y_i = p_i$ ($\vec{y} = \vec{p}$ in vector formulation).

Around this minimum, $F(\{y_i\})$ is Taylor expanded:

$$F(\vec{y}) = F(\vec{p}) + (\vec{y} - \vec{p}) \cdot \vec{\nabla} F(\vec{p}) + \frac{1}{2}(\vec{y} - \vec{p})^T \cdot \Delta F(\vec{p}) \cdot (\vec{y} - \vec{p})$$

where the gradient $\vec{\nabla} F(\vec{p})$ satisfies $(\vec{y} - \vec{p}) \cdot \vec{\nabla} F(\vec{p}) = 0$ and the Hessian matrix $\Delta F(\vec{p})$ is diagonal with $\Delta F(\vec{p})|_{ii} = 1/p_i$. Finally, we obtain:

$$P(\{N_i\} | M_k, \beta_k) \approx \frac{\sqrt{2\pi N}}{\prod_{i=1}^{N_b} \sqrt{2\pi N p_i(\beta_k)}} e^{-\frac{N}{2}\left[\sum_{i=1}^{N_b} \frac{(y_i - p_i(\beta_k))^2}{p_i(\beta_k)}\right]}.$$

Minimization of $\sum_{i=1}^{N_b} \frac{(y_i - p_i(\beta_k))^2}{p_i(\beta_k)}$ (generalized squared error) with respect to model parameters $\beta_k$ yields a parameter set $\beta_k^0$ that maximize the likelihood to observe the experimental JDD given the putative model $M_k$. In the main text, these parameters are called the *measured parameters*.

**Monte Carlo simulation of single molecules trajectories**

Planar single molecule trajectories were simulated using two-dimensional random walks. Free diffusion (model "D") was implemented by generating random, Gaussian distributed independent moves (steps) in the two directions of space at every observation time point (regularly spaced by $\delta t = 20 ms$, a typical time resolution for SMT experiments (54)), using the *randn* function in MATLAB. The only parameter of the model, the diffusion coefficient D, is related to the variance of the space steps by $\langle \delta x^2 \rangle = \langle \delta y^2 \rangle = 2D\delta t$.

Noisy directed motion (model "V") was implemented by generating, for each trajectory, random displacements encompassing a directed component at the velocity V along a random orientation $\theta$ $\delta x = V\delta t \cos\theta$ and $\delta y = V\delta t \sin\theta$, and a "noise" component, to account for fluctuations of the position around the directed track, modelled by Gaussian distributed displacements with variance $\langle \delta x^2 \rangle = \langle \delta y^2 \rangle = k_V$. Typically, $k_V$ reflects the contribution of the molecular noise and therefore scales as $2D\delta t$.

Anomalous diffusion (motion model "A") was implemented using a continuous time random walk with waiting times (CTRW, (56, 90)). Each move was being attributed a random time, sampled from a long-tailed distribution, which do not necessarily correspond to the observation times every $\delta t$. In this purpose, we calculated the cumulative distribution of waiting times (probability that the molecule moves before a certain date *t*), which varies from 0 at $t=0$ to 1 at $t=\infty$. Then, we generated a random number between 0 and 1, and by identifying this number to the cumulative distribution we obtained the date of the next move. The position of the molecule at the observation times was inferred from its position immediately before the next move. For the distribution of waiting times, we chose a distribution with a Pareto-like tail (decrease as $(1/t)^{\alpha+1}$ at $t >> \delta t'$), and a linear increase with time at short times $t < \delta t'$, with the regularization cutoff $\delta t' = \delta t / 2000 = 0.01 ms$. Similar to the free diffusion process, random space steps were generated from a Gaussian distribution of variance $\langle \delta x^2 \rangle = \langle \delta y^2 \rangle = 2D_\alpha (\delta t')^\alpha$. This procedure yields anomalous diffusion at time lags $\tau >> \delta t'$, with an anomalous exponent $\alpha$ and an effective diffusion coefficient $D_\alpha$ (56). Composite models DD, DV, DA were implemented by generating a fraction $f_D$ of free diffusion trajectories, and a fraction $1-f_D$ of trajectories of the D, V or A type respectively.

Simulations were performed in MATLAB (The Mathworks, Natick, MA), on a MacBook Pro laptop hosting a 2.53 GHz (P8700) Intel Core 2 Duo CPU. An overall working precision of $10^{-6}$ was chosen for numerical computations, and integrations over parameter spaces were performed using Simpson quadrature in MATLAB.

# Results and Discussion

**Characterizing single molecule transport using the Jump Distance Distribution**

The method described in this article was developed to analyze a set of $N$ individual planar trajectories, or fraction of trajectories, of identical duration $\tau$ (or equivalently, equal length), produced by a SMT experiment. For each trajectory, the jump distance is the Euclidian (geometrical) distance between the first and the last points of the trajectory (see Materials and Methods, M&M, and Fig. S1A). The jump distance distribution (JDD) is defined by classifying the $N$ jump distances in $N_b$ ordered bins of fixed size (Fig. S1B). The distribution is therefore defined by the set of integer numbers $\left\{ N_i, i=1..N_b, \sum_{i=1}^{N_b} N_i = N \right\}$ corresponding to the number of trajectories in each bin for the analyzed dataset.

To illustrate the power of the method, we have chosen to fit this experimental JDD with 6 two-dimensional motion models which correspond to established biological transport modes (10, 48, 57, 62, 91): a free diffusion model, denoted with the letter 'D', parametrized by the diffusion coefficient $D$; a noisy directed motion model, denoted with the letter 'V', parametrized by the velocity $V$ of the directed transport along linear tracks and the variance $k_V$ of the molecule position at each step around the track (5, 48); a model of anomalous subdiffusion, denoted with the letter 'A' and parametrized by the anomalous diffusion coefficient $D_\alpha$ and the power exponent $\alpha < 1$; and finally 3 composite models, encompassing a fraction $f_D$ of freely diffusing molecules and a fraction $1-f_D$ transported either by free diffusion with a larger diffusion coefficient $D_2$ (model DD), directed transport (model DV) or anomalous subdiffusion (model DA) ; these composite models are characterized by $f_D$ and the parameters needed for each submodel.

Mathematical closed form expressions for the probability for a molecule to jump a distance $(i-1/2)\delta r$ during the time $\tau$ have been derived for various motion models in the past (jump distance probability, see M&M), including free diffusion with or without a linear bias corresponding to directed motion, confined diffusion (73), anomalous subdiffusion (56, 87) or superdiffusion (88) including Levy flights (89). The theoretical jump distance distributions $\left\{ \tilde{N}_i^{D,V,A}, i=1..N_b \right\}$ are easily obtained by multiplying the probabilities by the number of molecules, and JDDs for composite models are linear combinations of $\left\{ \tilde{N}_i^{D,V,A} \right\}$ and the fraction $f_D$ of freely diffusing molecules (M&M). The schematics Fig. S1B shows that long tails in the JDD (e.g., longer than for pure Brownian motion, dashed black curve) can be the consequence of both a faster (free) diffusion (dark blue), directed transport (green), or anomalous diffusion (red). Generally speaking, anomalous subdiffusion tends to be characterized by a main peak shifted towards the small distances in the JDD, while the tail at large distances is enhanced and extended compared to free diffusion. In contrast, directed transport is characterized by a

Gaussian-like peak centred at $r = V\tau$, with a spreading reflecting the positional noise $k_V$ in the directed motion.

The observed JDD can therefore be fitted with multiple transport models. In practice, multiple models can provide satisfactorily fits of the experimental data, and a model selection procedure is required.

**Bayesian approach for model selection on JDD data**

For $K$ competing models $M_{k=1..K}$ (here $K=6$), and an observed JDD $\{N_i\}$, the probability of each model to be the actual transport mode that produced $\{N_i\}$ is given by the Bayes theorem (77), and reduces to:

$$P(M_k \mid \{N_i\}) = \frac{P(\{N_i\} \mid M_k)}{\sum_{k=1}^{K} P(\{N_i\} \mid M_k)}$$

when all plausible models are considered equiprobable. The marginal probability $P(\{N_i\} \mid M_k)$ is obtained by integrating $P(\{N_i\} \mid M_k, \beta_k) P(\beta_k \mid M_k)$ over the model parameter space (see M&M). $\beta_k$ denotes the entire set of parameters.

We first estimate $P(\{N_i\} \mid M_k, \beta_k)$. For statistically independent jump distances for different trajectories, the probability $P(\{N_i\} \mid M_k, \beta_k)$ to obtain the observed JDD given a model and a set of parameters is a multinomial distribution. Simplification of this distribution using the Stirling formula to express factorials, followed by a Laplace saddle-point approximation (see M&M for details) leads to:

$$P(\{N_i\} \mid M_k, \beta_k) \approx \frac{\sqrt{2\pi N}}{\prod_{i=1}^{N_b} \sqrt{2\pi N p_i(\beta_k)}} e^{-\frac{N}{2}\left[\sum_{i=1}^{N_b} \frac{(y_i - p_i(\beta_k))^2}{p_i(\beta_k)}\right]}$$

where $\{y_i = N_i / N\}$ is the experimental JDD normalized to the number of trajectories in the set, and the reference probabilities $\{p_i = \tilde{N}_i^{M_k} / N\}$ depend on the model and its parameters $\beta_k$.

Thus, for each model the parameters that maximize the likelihood of the observed configuration are those which minimize the generalized squared error $\sum_{i=1}^{N_b} \frac{(y_i - p_i(\beta_k))^2}{p_i(\beta_k)}$ where the squared error at each point is weighted by the inverse of its variance $w_i = 1/p_i$. To the leading (second) order in $y_i - p_i$ we can set $w_i = 1/y_i$, and the weights are then independent on model parameters. The minimization of $\sum_{i=1}^{N_b} w_i (y_i - p_i(\beta_k))^2$ with respect to model parameters is then performed using standard weighted least squares method based on the Newton-Gauss algorithm. The parameter set $\beta_k^0$ for which the generalized squared error is minimal can be interpreted as the *measured model parameters* (see (77) for details).

We next estimate $P(\beta_k | M_k)$. Owing to the stochastic nature of molecular transport, two finite samples collected from the same cell over the same period of time can be optimally fitted with slightly different parameters. Thus the *measured* $\beta_k^0$ and the "true" *in vivo* value might differ by some uncertainty $\delta\beta_k$, and model selection could be affected by the determination of optimal model parameters. Following the procedure described in (77), we used uniform distributions for $P(\beta_k | M_k)$ within a certain range $\beta_k^{\min} < \beta_k < \beta_k^{\max}$ and $P(\beta_k | M_k) = 0$ elsewhere; thus the integration over parameters $\beta_k$ represents a simple averaging of the peaked probability $P(\{N_i\} | M_k, \beta_k)$ over an integration range $[\beta_k^{\min} = \beta_k^0 - 10\delta\beta_k, \beta_k^{\max} = \beta_k^0 + 10\delta\beta_k]$ that includes all plausible values of $\beta_k$. This procedure provides an opportunity to penalize models with too many parameters, and also models for which parameters may be measured inaccurately.

To estimate both the accuracy of parameters measurements on which the method relies, and the uncertainties $\delta\beta_k$ that define the integration ranges, we have performed generalized squared error minimization on stochastic *in silico* trajectory datasets, simulated using Monte Carlo algorithms (see M&M).

**Performance of parameters measurement on simulated trajectories**

For each model and parameter set, a total of 20 trajectory sets were simulated (*in silico* experiments) with the (unless otherwise specified) default parameters:

- Trajectories and analysis parameters: $N = 3000, N_b = 30, \tau = 0.14s$
- Motion models parameters:
  $D = 0.02\mu m^2/s, D_2 = 0.1\mu m^2/s, V = 1.2\mu m/s, k_V = 0.0008\mu m^4/s^2,$
  $D_\alpha = 0.02\mu m^2/s^\alpha, \alpha = 0.5, f_D = 0.5$

Examples of simulated trajectories are shown on Fig. S1A. In order to focus on the efficiency of our method for short trajectories, we have chosen a default time lag that corresponds to trajectories of only 8 points. Then the generalized squared error between the JDDs produced by each *in silico* experiment and the tested models was minimized with respect to model parameters using a Gauss-Newton algorithm, with initial parameter values estimated automatically from the localization of the peak(s) in the JDD. The convergence of the algorithm was decided as soon as the relative change in squared error between two successive iterations was less than the working precision. This condition was not always achieved, owing to the oscillatory behaviour of some fitting functions and numerical uncertainties. In this case, a maximal number of iterations of 100000 (models D, V, DD, DV) or 5000 (models A, DA) was used instead, and convergence was visually checked.

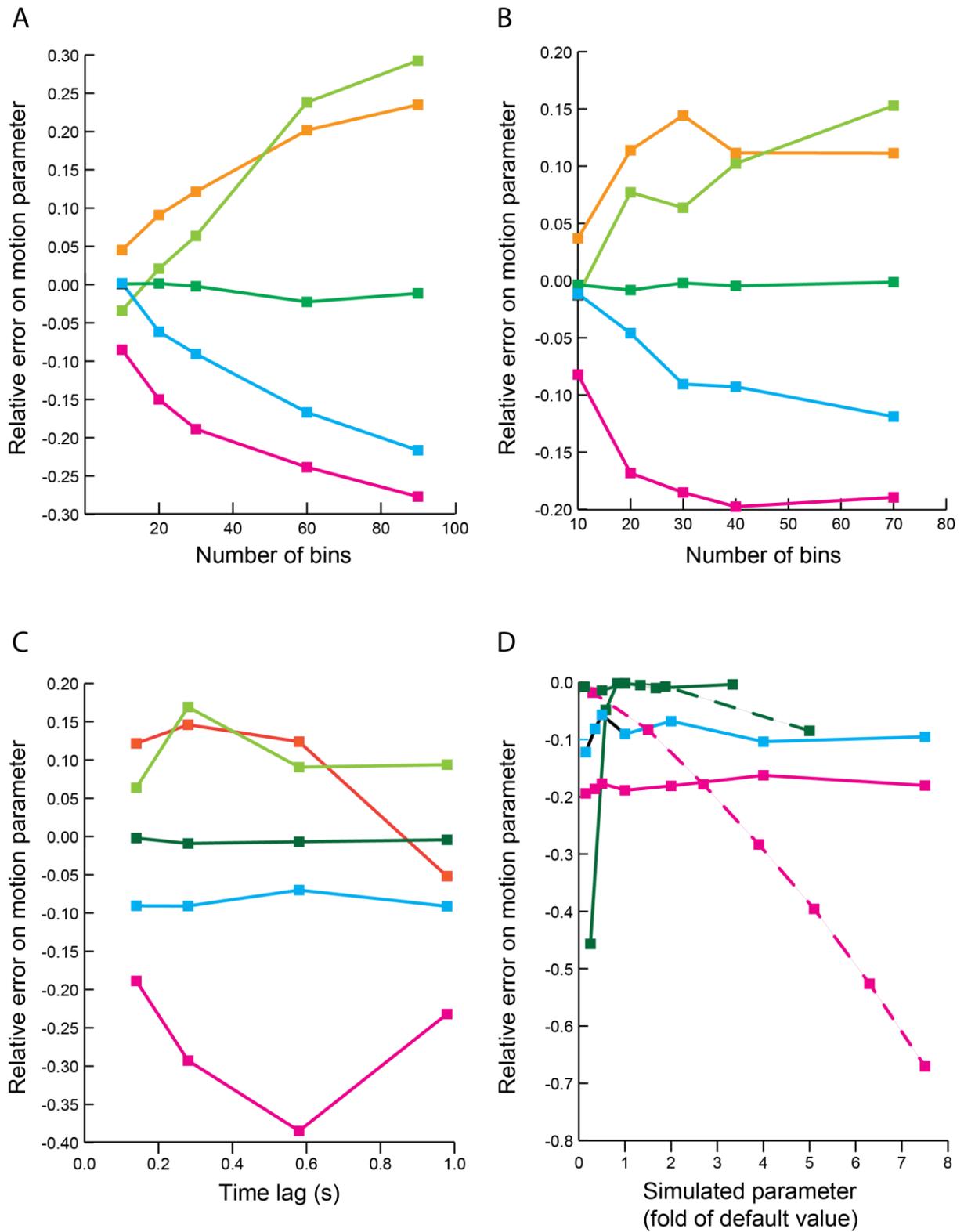

**Figure 1: Automated least squares fitting of the JDD reliably extracts molecular motion parameters.** Relative error on measured motion parameters as a function of: the number of bins with constant $N$ (A) and constant $N/N_b$ ratio (B), the time lag (C), and simulated motion parameters (D). Shown motion parameters are the diffusion coefficient $D$ for the D model (cyan), the transport velocity $V$ and positional variance $k_V$ for the V model (dark and light green respectively), and the anomalous diffusion coefficient $D_\alpha$ and power exponent $\alpha$ for the A model (magenta and orange, respectively). For panel (D), dark green and magenta dashed lines indicate the errors on estimated $V$ and $D_\alpha$ as a function of varying $k_V$ and $\alpha$ respectively, while plain lines represent the errors on estimated parameters for different input values of the same parameter. Simulated parameters are given in units of default parameters (see main text), except for $\alpha$ (dashed magenta line) where data points correspond to $\alpha = 0.3, 0.4, \ldots 0.9$, from left to right. Each data point was obtained after averaging over 20 simulations.

Results of these tests are displayed on Fig. 1, where the relative error to which model parameters are measured is shown as a function of $N_b$ with constant N (1A) and constant $N/N_b$ (1B), time lag (1C), and model parameters (1D), for the models D, V, and A. The relative error is defined as the difference between the parameter value used to simulate the trajectory set and the value measured through minimization of the error function, normalized to the former.

Fig. 1A shows that increasing the number of bins while keeping the number of trajectories constant decreases the accuracy of diffusion coefficient measures, both for D and A models. This is a consequence of the statistical noise which blurs the JDD when the condition $N/N_b \gg 1$ is not satisfied. In contrast, the velocity parameter is measured with a great accuracy over the entire range of bin numbers for the V model. Fig. 1B shows that, in contrast, increasing the number of bins while keeping the ratio $N/N_b$ constant was less affecting diffusion coefficient measures. This demonstrates that increasing the resolution of the JDD (number of bins) without reducing the accuracy of the fitting procedure can be done by simply increasing the number of trajectories, which is always possible, at least technically. In addition, Fig. 1C shows that, for the default parameters chosen in this study, the time lag, or equivalently here the number of trajectory points, does not influence the accuracy of parameter determination for the non-composite motion models. Finally, Fig. 1D shows that broad ranges of motion model parameters are measured with a typical accuracy of less than 10%, confirming that the JDD is an excellent measurable quantity to unveil transport features from short trajectories. It is noteworthy that the fitting accuracy is significantly decreased for very low directed transport velocities (green line) and anomalous diffusion with large power exponent $\alpha$ (dashed magenta line), two parameter regimes in which the A and V asymptotically tend towards free diffusion and lose their particular character. In contrast, the anomalous power exponent is measured more accurately in this large $\alpha$ regime (Fig. S2).

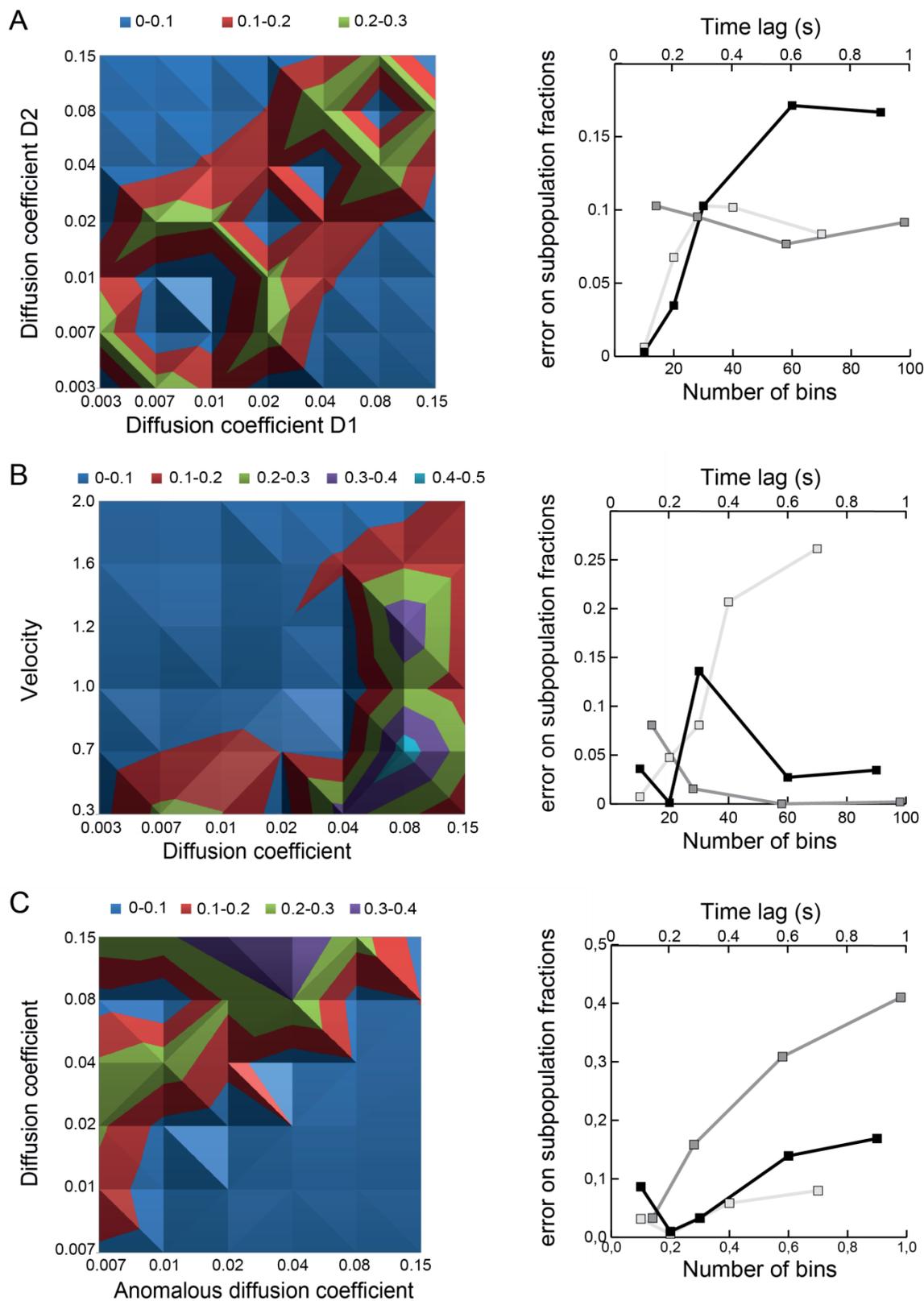

**Figure 2: Automated least squares fitting can separate contributions to the JDD of molecule subpopulations experiencing different motion modes.** Absolute error on the determination of the fraction $f_D$ of freely diffusing molecules for composite DD model (A), DV model (B) and DA model (C), with simulated $f_D = 0.5$. Color-coded diagrams (left) represent the error as a function of motion model input parameters (horizontal and vertical axis). Plots (right) represent the error as a function of $N_b$ with constant $N$ (black) and constant $N/N_b = 100$ (light grey), and of the time lag (dark grey). Each data point was obtained after averaging over 20 simulations.

Then, we estimated the accuracy to which subpopulations experiencing different modes of transport are identified. In this purpose, heterogeneous trajectory sets were simulated and their JDDs were fitted with theoretical JDDs for composite models (M&M). Results of these tests are displayed on Fig. 2 for the DD model (2A), DV model (2B) and DA model (2C). Fig. 2A shows that the contributions of two subpopulations of freely diffusing molecules with different diffusion coefficients $D_1$ and $D_2$ are distinguished with a small error (less than 0.2) on their relative fraction whenever $D_2 > 2 - 3D_1$ (left panel, red and blue). Therefore, in practical applications, it will be possible to distinguish organelle-bound proteins (with typical diffusion coefficient $D_1 = 0.001 - 0.05 \mu m^2/s$ (55, 63)) from cytosolic proteins (with typical diffusion coefficient $D_2 > 1 \mu m^2/s$). As for the simple models, increasing the bin number without increasing the number of trajectories is detrimental to the fitting accuracy (right panel, black line), while it does not significantly affect the error on subpopulation fractions when $N$ is also increased (light grey). Furthermore, modifying the time lag does not change the fitting accuracy in this situation (dark grey). It is noteworthy that this right panel was obtained in the default parameter regime where subpopulation fractions are accurately measured according to the color-coded diagram.

Similarly, Fig. 2B (left panel) shows that two subpopulations experiencing respectively free diffusion and (noisy) directed transport can be generally distinguished with a high accuracy (error on their relative fraction less than 0.2, red and blue regions of the diagram). According to these results, (slowly) diffusing membrane-associated proteins will be distinguished from actively transported proteins that move along directed tracks faster than $0.5 - 0.7 \mu m/s$, which is in the range of myosin motor-mediated transport (48, 92). Even fast diffusion with $D = 0.08 - 0.15 \mu m^2/s$, relevant in some biological contexts (10), will be distinguished from directed transport for $V > 1 - 1.5 \mu m/s$ (for our default analysis parameters), requirement that is also achieved *in vivo (92)*. In this DV scenario, increasing the time lag significantly improves the determination of subpopulation fractions (Fig. 2B). Indeed, the position of the diffusion and directed transport peaks in the JDD scale with $\sqrt{MSD(\tau)}$, respectively $\sqrt{\tau}$ and $\tau$. Thus, increasing the time lag tends to separate those peaks in the JDD, making the identification of subpopulations easier. Surprisingly, increasing the number of bins didn't improve the fitting accuracy (black, light grey).

Fig. 2C (left panel) shows that a composite population encompassing freely and anomalously diffusing molecules can be resolved if free and anomalous diffusion coefficients are significantly different (error on their relative fraction less than 0.2, red and blue regions). In contrast with the DD model though, where the "worst case scenario" is achieved when $D_1 \approx D_2$, the diffusion coefficients regime where D and A components are not well distinguished depends in practice on the anomalous power exponent $\alpha$. Here, for $\alpha = 0.5$ the fractions are well determined for $D < D_\alpha$, $D \approx D_\alpha$, and $D >> D_\alpha$ (see left of the diagram), but not for $D \geq D_\alpha$. Contrary to the DV model, increasing the time lag is detrimental to the fitting accuracy (see Fig. 2C, right, dark grey). As described for other models, increasing the number of bins tends to reduce the accuracy in the other parameter ranges (black line), less if the ratio $N/N_b$ is maintained (light grey line).

In summary, we have proven that automated least squares-based fitting of simulated JDD reliably extract the fraction of two subpopulations for the composite models DD, DV, and DA, over broad ranges of parameters. The JDD of a composite population might exhibit multiple distinct peaks depending on how different the subpopulations motion parameters are. While it might be required to increase the number of bins to resolve close peaks, this operation generally increases the noise of the JDD, and therefore reduces the fitting accuracy, unless more trajectories are considered to plot the JDD. Furthermore, while increasing the time lag allows separating free diffusion from directed transport, this can negatively impact the distinction of anomalous from free diffusion, indicating that an optimal time lag might also be required.

How well are motion parameters measured for the simulated composite models? For conciseness we didn't represent the relative error on measured parameter as a function of all other parameters as this would have yielded dozens of plots similar to Fig. 1. Instead, we generated scatter dot plots showing, for each composite model, the relative error on measured parameter as a function of the error on the estimated subpopulation fraction for the same simulated JDD (Fig. 3).

These plots show that motion parameters are generally reliably measured (relative error less than 0.2, vertical axis) when the subpopulations fractions are also accurately determined (error less than 0.2, horizontal axis). Thus, conclusions drawn on the ability of the least squares fitting to distinguish between subpopulation fractions (Fig. 2) generally extend to the estimation of underlying motion parameters. In contrast, the measurement of free diffusion coefficients, (Fig. 3A-C, light blue and Fig. 3A, dark blue), directed transport velocity (Fig. 3B, dark green) and anomalous diffusion coefficient (Fig. 3C, magenta) are generally inaccurate when $f_D$ is inaccurate. The positional variability parameter $k_V$ for the DV model (Fig. 3B, light green) and the anomalous power exponent for the DA model (Fig. 3C, orange) are more likely to carry larger errors despite an accurate determination of $f_D$. In agreement with the conclusions of (69), for uneven composite simulated subpopulations the motion parameters are always well determined for the dominant fraction, with a (small, <0.3) relative error that decreases with an increasing fraction of the corresponding subpopulation (Fig. S3).

Finally, the relative errors displayed on Figs. 1-3 were used to estimate a typical uncertainty $\delta\beta_k$ on the measurement of each parameter, yielding reduced integration ranges of the parameter space as defined in the previous section.

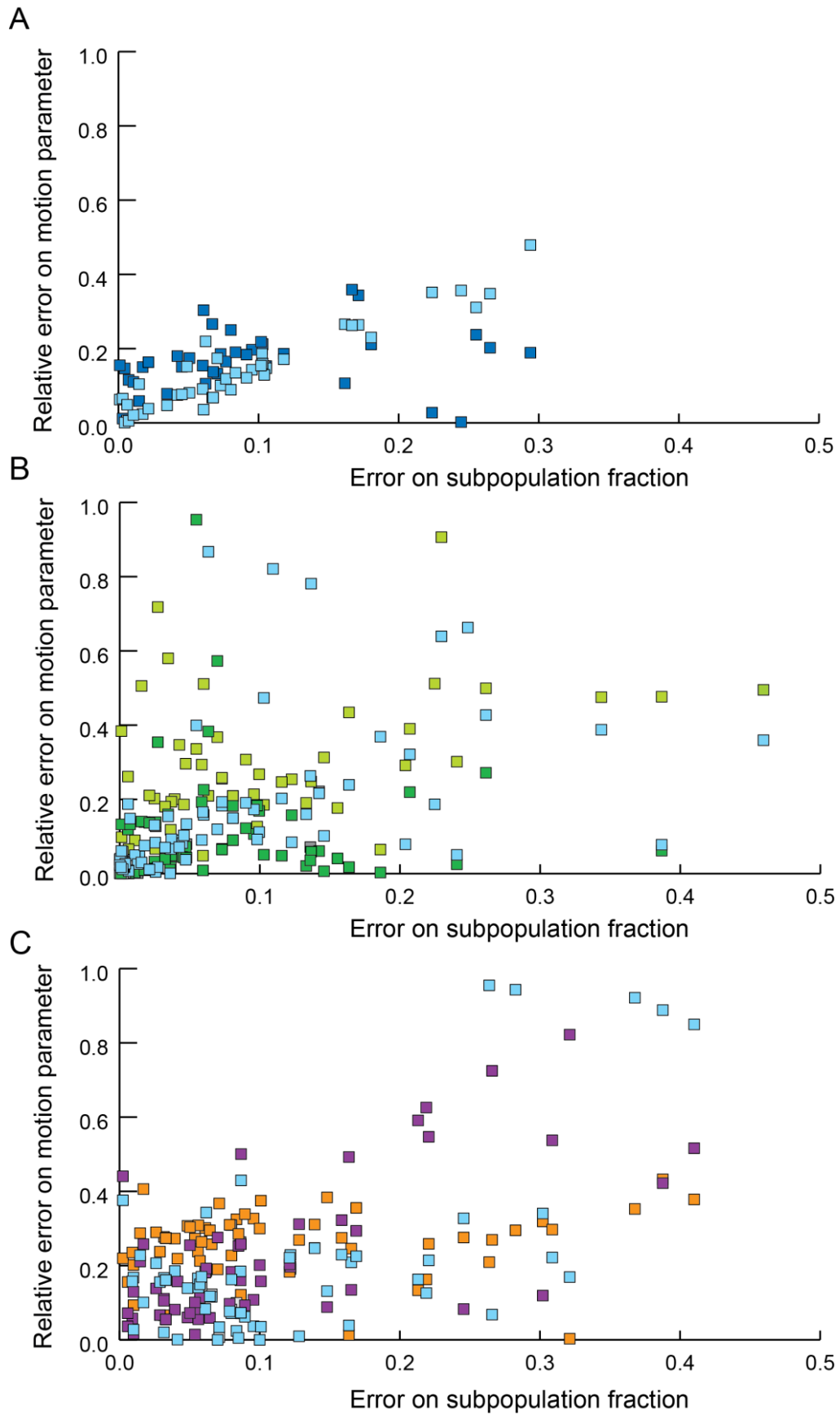

**Figure 3: Motion parameters are well measured when estimated subpopulation fractions are accurate.** Scatter dot plots showing, for all the motion model and trajectory parameters tested, the relative error on motion parameters measurements (y axis) as a function of the error on $f_D$ (x axis) for the DD (A), DV (B) and DA (C) models. Shown motion parameters are the diffusion coefficient of the freely diffusing fraction (cyan, A-C), the (fast) diffusion coefficient of the other freely diffusing fraction for the DD model (dark blue, A), the velocity and positional variance for the DV model (dark and light green respectively, B), and the anomalous diffusion coefficient and power exponent for the DA model (magenta and orange respectively, C). Each data point was obtained after averaging over 20 simulations.

**Performance of Bayesian model selection on simulated trajectories**

We next performed *in silico* tests of the reliability of the JDD-based Bayesian method for model selection (Figs. 4 and S4). For each motion model, and parameters representative of the different ranges determined in the previous section, 10 JDDs were simulated and analyzed with the complete Bayesian model selection procedure (as described in M&M), yielding for each simulated JDD a probability that each tested model is « true ». When no model got a probability larger than 75%, the run was counted as « undetermined ». The output of this analysis was represented as bar charts showing the occurrence, over 10 runs, of determined motion models for each input model and parameter range.

Fig. 4A shows that irrespective of the diffusion coefficient, free diffusion is rarely mistaken with other types of transport (first two bars from left to right). In the presence of low to moderate noise, pure V model is also well determined (third and fourth bars), although it might be, in the small velocity regime, interpreted as a DV with a very small freely diffusing subpopulation that compares with the typical uncertainty on $f_D$ (Fig. 2B left panel). Thus, after qualitative inspection of the result, the analyst still recognizes a V model. Similarly, significant subdiffusive character (small $\alpha$) is accurately extracted, though a simple A model might be mistaken with a DA model with $f_D$ comparable to the typical error on fraction measurement. Here again, qualitative inspection of the result is sufficient to recover the actual nature of the underlying motion (A model). Our Bayesian model selection procedure is even more efficient when applied to composite JDDs (Fig. 4B). Indeed, in the parameter regimes where $f_D$ and motion parameters are well determined by the automated least square fitting, the right model is determined in almost 100% of the test runs. Significant deviations between the simulated and estimated models are noticed only in parameter regimes where parameter measurement is inaccurate (Fig. S4). In some cases, increasing the number of bins (and the number of trajectories) significantly improves motion model determination (V model with large *V* and high noise, A model with large $D_\alpha$ large $\alpha$); in contrast, in other cases (V model with small *V* and high noise, A model with small $D_\alpha$ large $\alpha$) there is an apparent strong contribution of free diffusion that do reflect the fact that slow directed transport might be drowned in high noise (Fig. S4A, small V/high noise and S3B, large gap high noise), while a weak anomalous character $\alpha \approx 1$ might be underestimated (Fig. S4A, small $D_\alpha$ large $\alpha$), or even missed (see model DA, large gap large $\alpha$ on Fig. S4B).

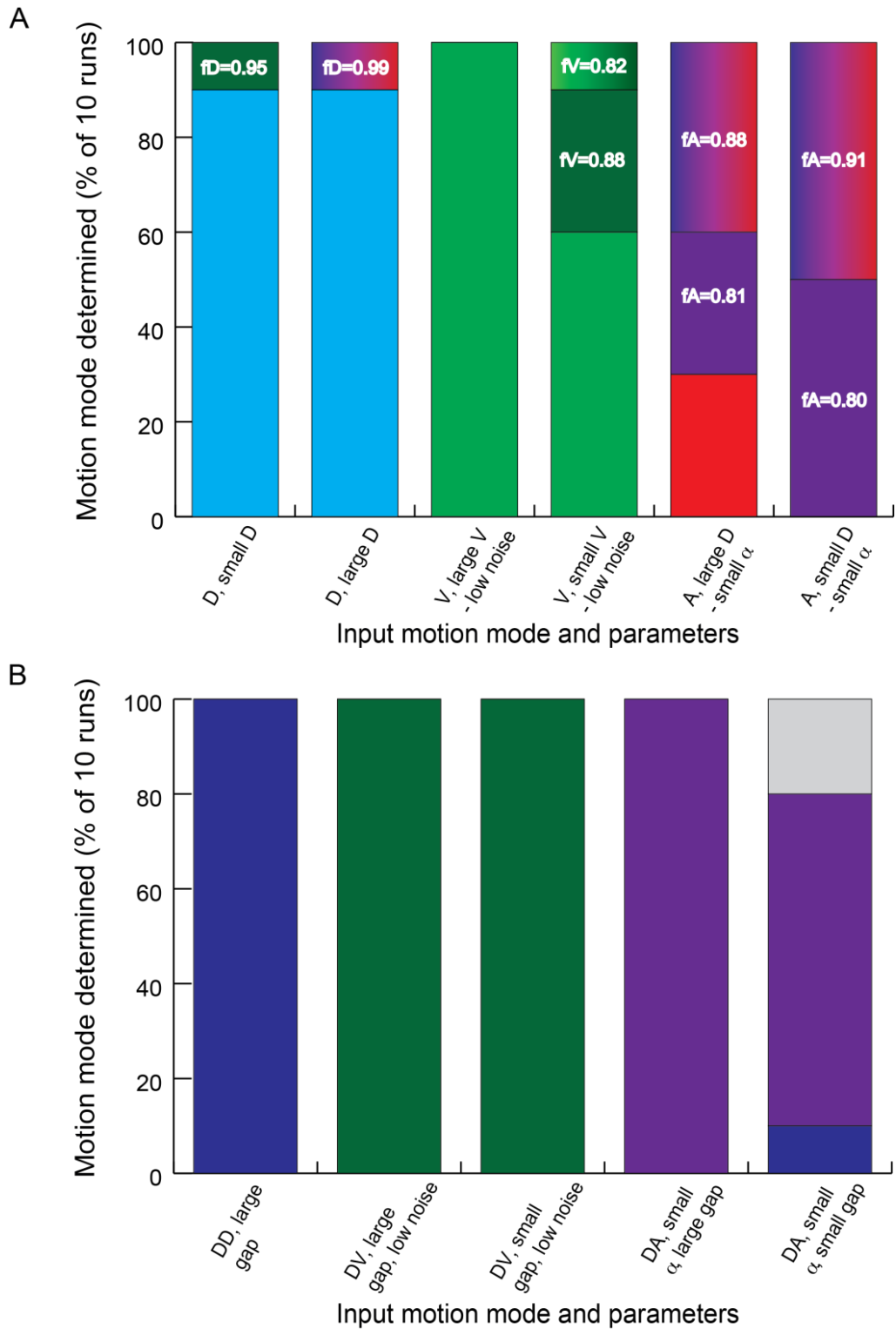

**Figure 4: A Bayesian model selection procedure extracts underlying transport mode from simulated JDD.** Bar charts showing the occurrence of model determination for various input (simulated) models and parameter ranges. Simple and composite input motion modes are represented on panels (A) and (B) respectively. Output motion modes are: D (cyan), V (light green), A (red), DD (dark blue), DV (dark green), DA (purple), and undetermined (grey). Note that special undetermined situations, where the probabilities of V and DV (or A and DA) models almost sum to 1, were represented with light to dark green and magenta to purple colour gradients respectively for illustration purposes. For simple input models for which some runs showed a significant probability (larger than 25%) for the corresponding composite model, the average fraction of the simple input model in the composite output model was averaged over those runs and is shown in white. Those fractions are always close to 1, indicating that the contribution of the input motion model dominates in the detected composite model.

# Conclusion

SMT experiments should offer the possibility to probe the local environment of molecules of interest, and thus to gain considerable insight on their individual motion, interactions, and overall intracellular dynamics. For these reasons, they are increasingly used in a variety of biological contexts (11, 13, 16, 62, 63, 93, 94). However, despite recent progress in fluorophores engineering (95, 96) and tracking algorithms, it is still difficult to follow single molecules for a long time while keeping photo-bleaching, trajectories merging or clustering to a reasonably low level, especially when tracked molecules are clustered in small subcellular regions. Hence, the trajectories obtained are usually short, and in this limit current analysis methods fail to accurately determine motion features (36, 45, 46, 60). In addition, we still lack a method suitable to discriminate between multiple transport modes experienced by a single molecule along an individual trajectory.

In this work, we have derived a new method to analyze quantitatively SMT data and extract key motion features and parameters. The method was carefully derived mathematically from rigorous statistical considerations on a variable directly measurable from individual trajectories, the Jump Distance (69). We show than combining a generalized least squares fitting of the JDD by particular motion models with a Bayesian model selection procedure offers a good alternative to MSD-based analysis, that is also efficient for short trajectories, where the latter fails (36). After having derived analytically the mathematical foundations of the method, we performed generalized least square fitting of simulated JDDs corresponding to biologically relevant transport modes over a broad range of motion parameter. We showed that in most parameter regimes, measurements of the input motion parameter values were accurate, with typically less than 20% error using realistic trajectories length and number. The fitting was efficient both for homogeneous or heterogeneous trajectory sets (Figs. 1-3 and S2-S3). In both cases, the Bayesian model selection procedure was able to determine the underlying transport mode(s) from almost all simulated JDDs (Fig. 4). Note that parameters relative to anomalous diffusion were overall measured with less accuracy than for other motion modes. This is probably due to an incomplete sampling of the waiting time distributions when simulating Continuous Time Random Walks with finite time steps, leading to alterations in the low distance end of the simulated JDD compared to the ideal theoretical JDD. Improving the CTRW algorithm should strengthen the tests of our analysis method.

In contrast, parameter regimes in which the subpopulation fractions (Figs. 1-3) and underlying transport modes (Fig. S4) of heterogeneous trajectory sets were less accurately extracted were also identified. For clarity and conciseness, we could not report tests of all possible parameter values; however, our analysis was sufficient to extract a "recipe" on how to use our JDD-based Bayesian method to analyze real samples, depending on *a priori* information available on molecular motion.

The JDD of a uniform population has a peak whose position scales as $\sqrt{MSD(\tau)}$, where $\tau$ is the (chosen) time lag, e.g. as $\sqrt{D\tau}$, $\sqrt{D_\alpha \tau^\alpha}$ and $V\tau$ for free, anomalous diffusion, and directed transport respectively (77). Distinguishing directed transport from free or anomalous diffusion will require larger time lags, while distinguishing free from anomalous diffusion will necessitate either large or short time lags (regimes where the functions $\sqrt{D\tau}$ and $\sqrt{D_\alpha \tau^\alpha}$ diverge from each other), in order to resolve the peaks in the JDD. Tuning the time lag can be achieved both by choosing the frequency of imaging, and by selecting the adequate number of points in the trajectories (or fractions of split trajectories). Once the time lag is chosen, the obtained JDD extends over a given range of jump distances: the choice of the number of bins depends then on this range and on the resolution needed to resolve the different peaks. Finally, the size N of the sample has to be at least 100 times the number of bins to prevent statistical noise in the JDD.

Our methodology can be straightforwardly applied to a broad range of systems, including in vitro experiments. In addition, its analytical formulation can be formally extended without new derivation to include other transport modes with a known expression for the jump distance probability distribution, like for instance confined diffusion (73), fractional diffusion (87) or superdiffusion (88) including Levy flights (89). We have restricted our numerical tests to D, V, and A modes only for conciseness, and because of their relevance for proteins, or lipid vesicles, moving in heterogeneous media (97, 98) and their established contribution to cellular dynamics (10, 48, 57, 62, 91). Strickly speaking the method requires that jump distances of individual trajectories are uncorrelated. For instance, interacting proteins simultaneously visualized would not satisfy this requirement. However, the use of switchable fluorophores (like the mEOS) for which a very small number of molecules are simultaneously visualized limits the occurrence of correlated trajectories. Correlations might also appear when short trajectories are obtained from the splitting of non-markovian trajectories. In this case, correlations extend over a certain distance along the trajectory, and therefore increasing the number of subtrajectory points (e.g. the time lag) above this correlation length should reduce correlation effects.

Finally, the method can be readily extended to include composite populations encompassing 3 or more motion modes; this would require adding new terms to the linear combination defining jump distance distributions for composite models, with new subpopulation fractions as additional parameters. We are currently working on the creation of a graphical interface that will allow the user to choose the motion modes to include, the number of subpopulations to search for and the expected ranges of motion parameters to make the method more user friendly.

## Acknowledgements

We thank Derek McCusker for initiating this project and careful reading of the manuscript. We also thank Mike Tyers and Eric Schirmer for advice and complementary funding. This work was funded in France by the Agence pour la Recherche sur le Cancer (ARC grant PDF20120605172) and CNRS, and in the UK by the Wellcome Trust Centre for Cell Biology and the University of Edinburgh. The author declares no conflict of interest.

# Supporting Material

Four supplementary figures are provided below.

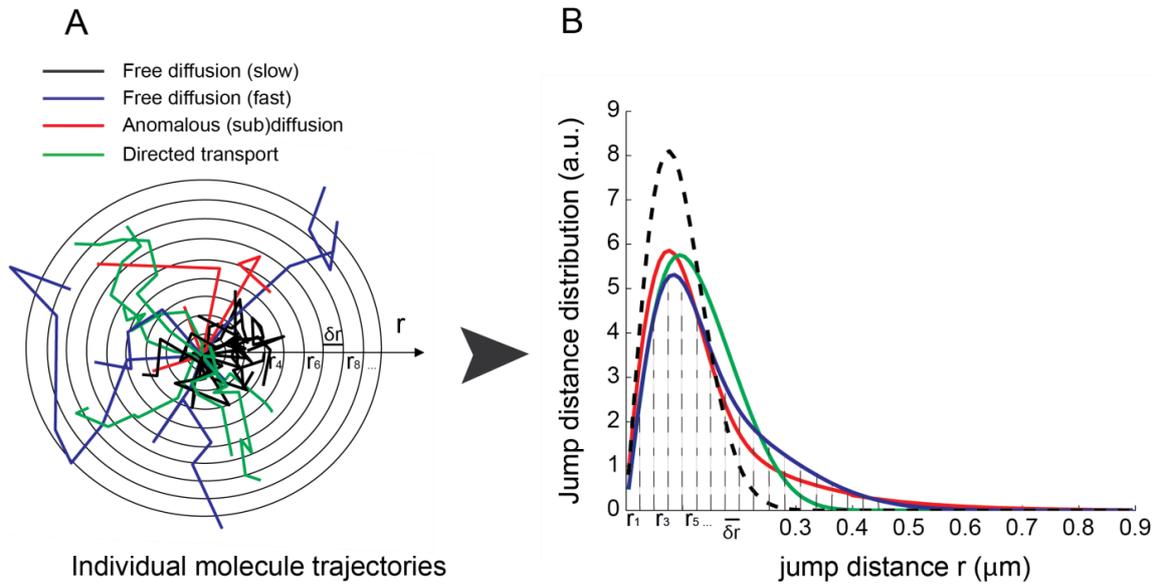

**Supplementary Figure S1: Construction of the Jump Distance Distribution.** (A) Examples of planar single molecule trajectories of equal length/duration, gathered to the same starting point. Shown are trajectories from simulated slow (black) and fast (blue) free diffusion, anomalous subdiffusion (red), and noisy directed motion (green). From the starting point, circular bins of size $\delta r$ are plotted, and trajectories are classified in the bins according to the position of their ending point, yielding the Jump Distance Distribution (see M&M). (B) Examples of JDDs obtained for a homogeneous population of freely diffusing molecules (black), and composite populations sharing long-tailed JDDs including also anomalously diffusing molecules (red), fast freely diffusing molecules (blue), and actively transported molecules (green). These three transport modes provide however very distinctive features to the JDD.

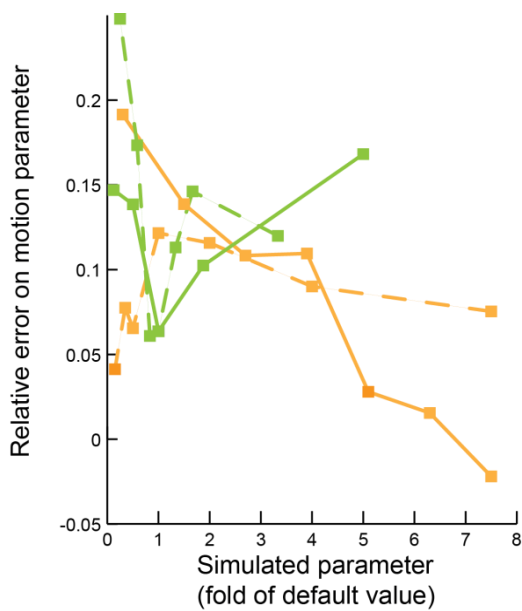

**Supplementary Figure S2: Measurement of additional motion parameters from automated least squares fitting of the JDD.**

Relative error on measured motion parameters as a function of simulated motion parameters. Shown measured parameters are the positional variance $k_V$ for the V model (light green), and the anomalous power exponent $\alpha$ for the A model (orange). Light green and orange dashed lines indicate the errors on estimated $k_V$ and $\alpha$ as a function of varying $V$ and $D_\alpha$ respectively, while plain lines represent the errors on estimated parameters for different input values of the same parameter. Simulated parameters are given in units of default parameters (see main text), except for $\alpha$ (plain orange line) where data points correspond to $\alpha = 0.3, 0.4, 0.5 \ldots 0.9$, from left to right. Each data point was obtained after averaging over 20 simulations.

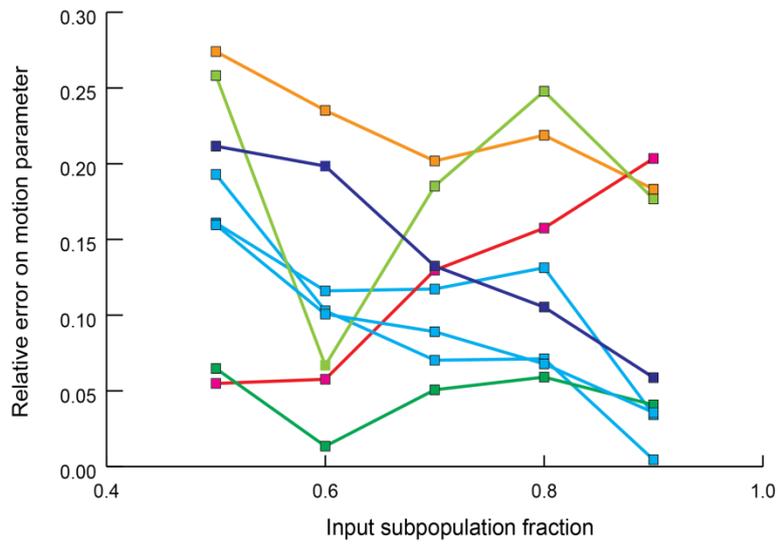

**Supplementary Figure S3: The accuracy of motion parameter measurement increases with the subpopulation fraction for composite input models.**

Relative error on measured motion parameters as a function of the input subpopulation fraction for the composite models. Shown motion parameters are the diffusion coefficient of the (slowly) freely diffusing fraction (cyan, 3 lines corresponding to DD, DV and DA input models), the (fast) diffusion coefficient of the other freely diffusing fraction for the DD model (dark blue), the velocity V and positional variance $k_V$ for the DV model (dark and light green respectively, B), and the anomalous diffusion coefficient $D_\alpha$ and power exponent $\alpha$ for the DA model (magenta and orange respectively, C).

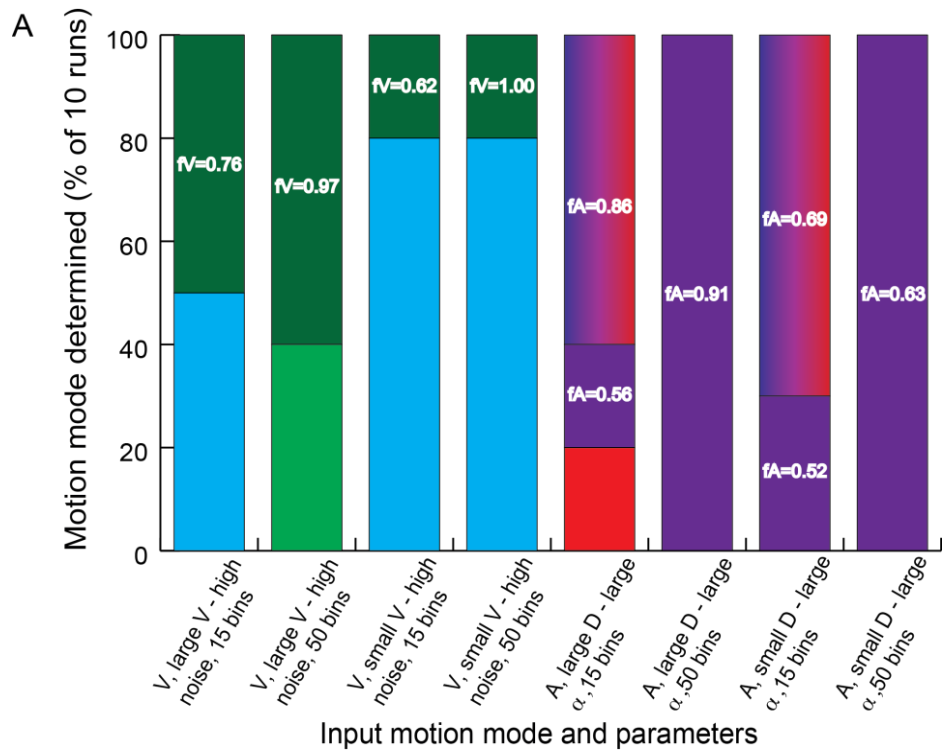
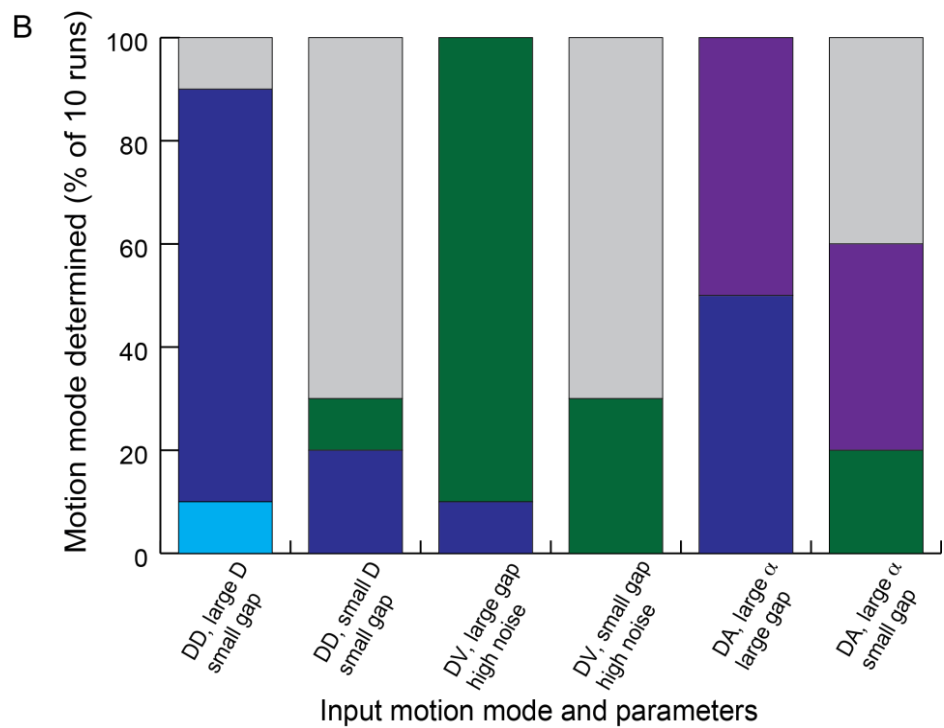

**Supplementary Figure S4: Limits of the Bayesian model selection procedure and partial improvement with increasing number of trajectories.**

Bar charts showing the output of the Bayesian model selection procedure for various input (simulated) motion models in parameter ranges for which the least squares fitting is less accurate. Simple and composite input motion modes are represented on panels (A) and (B) respectively. The colour coding of possible outputs and the simple model fractions in composite and undetermined output models in Panel A are defined as in Fig. 4.